\documentclass[prl,twocolumn,showpacs]{revtex4}
\usepackage{graphicx}

\begin{document}

\title{Dynamics of an Ensemble of Noisy Bistable Elements with 
Global Time-Delayed Coupling}
\date{\today}
\author{D. Huber and L. S. Tsimring}
\affiliation{Institute for Nonlinear Science, University of California, 
San Diego, La Jolla, CA 92093-0402}

\begin{abstract}%
The dynamics of an ensemble of bistable elements with global
time-delayed coupling under the influence of noise is studied
analytically and numerically. Depending on the noise level the system
undergoes ordering transitions and demonstrates multi-stability. That
is, for a strong enough positive feedback it exhibits a non-zero
stationary mean field and a variety of stable oscillatory mean field
states are accessible for positive and negative feedback. The
regularity of the oscillatory states is maximal for a certain noise
level, i.e., the system demonstrates coherence resonance. While away
from the transition points the system dynamics is well described by a
Gaussian approximation, near the bifurcation points a description in
terms of a dichotomous theory is more adequate.
\end{abstract}

\maketitle

Noise induced hopping events in bi- or multi-stable systems form the
basis of many interesting phenomena observed in physics, biology,
chemistry, as well as social science. The study of such rate processes
has thus been a subject of great interest and various techniques such
as Langevin, Fokker-Planck and master equations have been used to
describe the dynamics of stochastic systems
\citep{Haenggi90,Desai78,Jung92}.

In this article we generalize a well-studied stochastic model
consisting of an ensemble of interconnected noise driven bistable
oscillators by introducing time-delayed couplings. Such an extension
is important since it has been realized that time-delays are
ubiquitous in nature and often change fundamentally the dynamics of
the system
\citep{Bresseloff98,Nakamura94,Choi85}.

The dynamics of the network is studied numerically by using Langevin
equations and analytically by two complementary mean field
descriptions which are derived from the corresponding Fokker-Planck
and master-equations, respectively.
 
We assume that the bistable elements are highly interconnected, so
that the connectivity can be approximated by a global all-to-all
coupling.  In such a system a bistable element may for instance model
the basic properties of a neuron that can either be in a firing or a
non-firing state, a person that can opt between two choices by means
of individual voting, or a gene that is either expressed or
non-expressed. The globally coupled system may then represent a highly
interconnected neural network \citep{Sompolinsky91}, a social group in
which the individual voting behavior is influenced by opinion polls
\citep{Zanette97}, or a genetic regulatory network
\citep{Gardner00}, respectively. Other examples are catalytic 
surface reactions \citep{Rose96} and allosteric enzymic reactions
\citep{Hess96}.

The properties of globally coupled elements have been a subject of
many studies \citep{Desai78,Jung92,Kuramoto91,Yeung99}. In particular,
\citet{Desai78} studied the synchronization of thermally activated
bistable elements with instantaneous coupling and found an exact mean
field solution in the thermodynamic limit $N\to\infty$, where $N$ is
the number of elements in the network. This system exhibits a second
order phase transition to an ordered state with non-zero stationary
mean field. The effect of a time-delayed coupling has been studied by
\citet{Yeung99} for a globally coupled network of periodic phase
oscillators and \citet{Tsimring01} investigated the dynamics of a
single bistable element driven by noise and time-delayed feedback.

Here, as in \cite{Yeung99}, it is assumed that the time delays between
the bistable elements are uniform. Such an approximation is
justified in certain neural networks, whose time-delays 
are remarkably constant, as suggested by recent findings
\cite{Salami03}. Similarly, for certain regulatory
genetic networks the response time lag is determined by a single time
constant \citep{Paulsson01}.

Our prototype system for the study of rate processes in extended
systems consists of $N$ Langevin equations, each describing the
overdamped noise driven motion of a particle in a bistable potential
$V=-x^2/2+x^4/4$, whose symmetry is distorted by a global coupling to
the time-delayed mean field $X(t-\tau)=N^{-1}\sum_{i=1}^Nx_i(t-\tau)$,
\begin{equation}
\dot{x}_i(t)= x_i(t)-x_i(t)^3+\varepsilon X(t-\tau)+\sqrt{2D}\xi(t),
\label{lang}
\end{equation}
where $\tau$ is the time delay, $\varepsilon$ is the coupling strength 
of the feedback
and $D$ denotes the variance of the Gaussian fluctuations $\xi(t)$,
which are mutually independent and uncorrelated
$\langle \xi_i(t)\xi_j(t') \rangle = \delta_{ij}\delta(t-t')$.

We first study system (\ref{lang}) numerically.  The simulations are
carried out using a fixed-step fifth-order Runge-Kutta method with
linear interpolation for the evaluations at intermediate steps
required for the delayed variables.  If not otherwise stated, the
time-step and the number of elements is $\Delta t=0.01-0.05$ and
$N=2500$, respectively.

For $\varepsilon=0$, the elements are decoupled from each other. They 
jump from one potential well to the other randomly and independently of
each other. Therefore, in this case the mean field $X=0$. 
For small $|\varepsilon|$, the mean field remains
zero. At a certain $\varepsilon=\varepsilon_{\rm st}>0$ which depends on
the noise intensity $D$, but is independent of the time delay, the
system undergoes a second order (continuous) phase transition and
adopts a non-zero stationary mean field.

For a negative feedback, a transition to a periodically oscillating
mean field solution is observed at a certain
$\varepsilon=\varepsilon_{\rm osc-}<0$. Here and for the rest of this
paper a $(-/+)$ index means that the corresponding value is associated
with a negative/positive feedback.

Above a certain noise level $D_{H}$ the transition at
$\varepsilon_{\rm osc-}$ is second order as well. However, for
$D<D_{H}$ the system exhibits a first order (discontinuous) transition
associated with hysteretic behavior. The critical noise strength $D_{H}$
depends on the time delay and is $D_{H}=0.07$ for $\tau=100$.

For large time delays $\tau\gg\tau_K$ ($\tau_K$ is the inverse Kramers
escape rate from one well into the other), depending on the initial
state the system adopts one of the many accessible oscillatory states
featuring different periods.  Even for a positive feedback, besides
the stationary solution several oscillatory states with periods
$T\lesssim\tau$ are observed for $\varepsilon>\varepsilon_{\rm
osc+}\gtrsim\varepsilon_{\rm st}$.  If the feedback is negative, the
system only has oscillatory non-trivial solutions. The observed
periods are $T\lesssim2\tau$ for $\varepsilon<\varepsilon_{\rm osc-}$.

The simulations show that for a negative feedback all oscillating
states have a vanishing time-averaged mean field $\langle
X\rangle_t$. However, for a positive feedback besides
symmetric periodic solutions with $\langle X\rangle_t=0$, states with
significant non-zero temporal mean are possible.

In order to theoretically study the dynamical properties of a globally
coupled set of noisy bistable elements (with no time delay),
\citet{Desai78} derived a hierarchy of equations for the cumulant
moments of the distribution function from the multi-dimensional
Fokker-Planck equation for the joint probability distribution for all
elements. For large noise intensities, when the statistics of
individual elements are approximately Gaussian, this hierarchy can be
truncated. Applying this approach to our system yields the following
set of equations for the mean field $X$ and the variance
$M=N^{-1}\sum(x_i-X)^2$,
\parbox{7.5cm}{
\begin{eqnarray*}
\dot{X}(t)&=& X(t)-X^3(t)-3X(t)M(t)+\varepsilon X(t-\tau),
\label{XM}\\
\frac{1}{2}\dot{M}(t)&=& M(t)-3X^2(t)M(t)-3M^2(t)+D.
\end{eqnarray*}
}\hfill
\parbox{8mm}{\begin{eqnarray}\end{eqnarray}}

To compare the predictions of the Gaussian approximation
(\ref{XM}) with the original Langevin model (\ref{lang})
we calculate the power of the main peak $P_{\rm peak}$ in the
power spectrum of $X(t)$. It is proportional to the amplitude of
the mean field oscillations and can thus be used to analyze the Hopf
bifurcation, which describes the transition to the oscillating
mean-field states. The pitchfork bifurcation
to the non-zero stationary mean field for a large enough
positive coupling is characterized by the dependence of the temporal
mean of the mean field $\langle X \rangle_t$ on the system parameters.

\begin{figure}
\includegraphics[width=8cm]{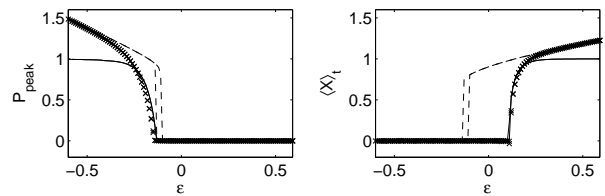}
\caption{\label{mixall} The peak power $P_{\rm peak}$ and $\langle
X \rangle_t$ as a function of the coupling strength for the Langevin
model (crosses), the Gaussian approximation (dashed line), where the
double line indicates hysteretic behavior, and the dichotomous theory
(solid line). The constant noise strength is D=0.1 and the time delay
is $\tau=100$. For For $X=0$ and $D=0.1$ the Kramers time is
$\tau_K=54.1$.}
\end{figure}

\begin{figure}
\includegraphics[width=8cm]{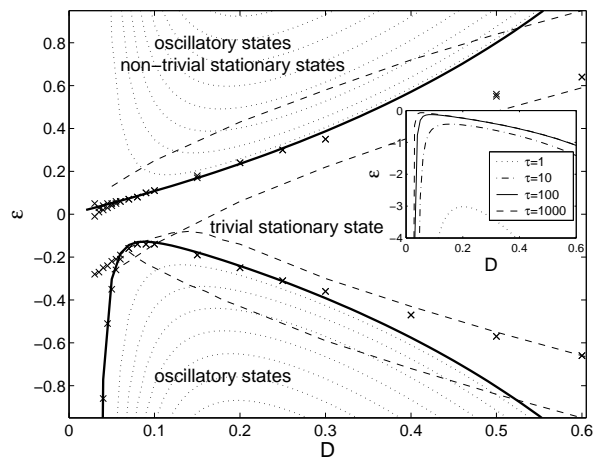}
\caption{\label{phase} Phase diagram for $\tau=100$ of the 
Langevin model (crosses), the Gaussian approximation (dashed lines)
and the dichotomous theory (solid lines and dotted lines).  The solid
line and the dotted line respectively depict the primary solution and
the higher order solutions of Eq. (\ref{omegaeps1}) and
(\ref{omegaeps2}). Phases separated by double lines indicate
hysteretic behavior. The inset shows how the Hopf bifurcation line
$\varepsilon^1_{\rm osc-}(D)$ varies with the time delay $\tau$. For
$X=0$ and $D<0.3$ the Kramers time is $\tau_K>10$. }
\end{figure}

For $\tau=100$ the peak power $P_{\rm peak}$ and the temporal mean
$\langle X \rangle_t$ resulting from the Gaussian approximation and
the Langevin model are shown in Fig. \ref{mixall} as a function of the
coupling strength $\varepsilon$. The phase diagrams of these models
are shown in Fig. \ref{phase} in the $(D,\varepsilon)$-parameter
plane.

Fig. \ref{mixall} shows that away from the bifurcation points the
Gaussian approximation describes the Langevin dynamics correctly.
However, near the transition points the Langevin dynamics is strongly
non-Gaussian even for large noise temperatures.  For instance, the
Gaussian approximation predicts that both bifurcations are subcritical
for the entire noise range $D=0.03-1.0$ considered in this study (see
Fig. \ref{phase}), while in the original Langevin model the bifurcations
are subcritical only for $D<0.07$.

Including higher-order cumulant equations only leads to a very slow
convergence towards the true solution of the Langevin model. Thus, in
order to describe the behavior of the system near the bifurcation
points, we apply a complementary {\em dichotomous} approximation,
which is valid in the limit of small noise, where the characteristic
Kramers transition time $\tau_K\gg 1$. In the dichotomous
approximation intra-well fluctuations of $x_i$ are neglected. Thus
each bistable element can be replaced by a discrete two-state system
which can only take the values $x_{1,2}=\pm1$. Then the collective
dynamics of the entire system is described by the master equations for
the occupation probabilities of these states $n_{1,2}$. This approach
has been successfully used in studies of stochastic resonance and
coherence resonance
\citep[e.g.][]{McNamara89,Gammaitoni98,Jung92,Tsimring01}. 
For instance, using this approach \citet{Jung92} found stationary mean
field solutions in a globally coupled, time delay free network of
bistable elements.

The dynamics of a single element is determined by the hopping
rates $p_{12}$ and $p_{21}$, i.e., by the probabilities to hop over
the potential barrier from $x_1$ to $x_2$ and from $x_2$ to $x_1$,
respectively.  In a globally coupled
system, $n_{1,2}$ and $p_{12,21}$ are identical for all elements.
Then the master equations for the occupation probabilities read
\begin{equation}
\label{ndot1}
\dot{n}_1  =  -p_{12}n_1+p_{21}n_2\ ,\,\
\dot{n}_2  =  p_{12}n_1-p_{21}n_2.
\label{ndot2}
\end{equation}

In the dichotomous approximation
the mean field  $X=x_1 n_1+x_2 n_2=n_2-n_1$, and making use of the
probability conservation $n_1+n_2=1$ we obtain the equation for the
mean field
\begin{equation}
\dot X(t) = p_{12}-p_{21}-(p_{21}+p_{12})X(t).
\label{XX1}
\end{equation}

The hopping probabilities $p_{12,21}$ are given by the Kramers
transition rate \citep{Kramers40} for the instantaneous potential
well, which in the limit of small noise $D$ and coupling strength
$\varepsilon$ reads \citep[cf.][]{Tsimring01},
\begin{equation}
p_{12,21}=\frac{\sqrt{2\mp 3\varepsilon X(t-\tau)}}{2\pi}
\exp\left(-\frac{1\mp4\varepsilon X(t-\tau)}{4 D}\right). 
\label{p1221}
\end{equation}

A linear stability analysis of Eq. (\ref{XX1}) near the trivial state
$X=0$ yields the transcendent equation for the complex eigenvalue
$\lambda$,
\begin{equation}
\lambda=\frac{\sqrt{2}}{\pi}{\rm e}^{-1/4D}
\left(\frac{\varepsilon(4-3D)}{4D}{\rm e}^{-\lambda\tau}-1\right).
\label{lambda}
\end{equation} 
For a positive coupling this equation always has a real solution.
At a certain critical coupling $\varepsilon_{\rm st}={4D}/(4-3D)$
the eigenvalue becomes positive indicating the pitchfork bifurcation
observed in the Langevin system (\ref{lang}).  Besides this real
solution, Eq. (\ref{lambda}) possesses an infinite number of complex
solutions corresponding to oscillating mean fields. However, only a
finite number of them corresponds to unstable modes at finite $\tau$
and $\varepsilon$ as we will see below. The critical values
$\varepsilon$ of the corresponding instabilities are found by
substituting $\lambda=\mu+{\rm i}\omega$ into Eq. (\ref{lambda}),
separating real and imaginary part and setting $\mu=0$:
\begin{eqnarray}
\label{omegaeps1}
\omega\tau &=&-\frac{\sqrt{2}}{\pi}\exp(-1/4D)\tau\tan\omega\tau\\
\label{omegaeps2}
\varepsilon_{\rm osc}&=&\frac{\varepsilon_{\rm st}}{\cos\omega\tau}.
\end{eqnarray}
This set of equations has a multiplicity of solutions, indicating that
multi-stability occurs in the globally coupled system beyond a certain
coupling strength. For finite time delays and positive
coupling, besides the stationary solution, several oscillatory states
with periods $T_k$ close to but slightly larger than $\tau/k$ are
observed for $\varepsilon>\varepsilon^{k}_{\rm
osc+}\;\;\{k=1,2,\ldots\}$, where the transition points are ordered as
follows, $0<\varepsilon_{\rm st}<\varepsilon^{1}_{\rm
osc+}<\varepsilon^{2}_{\rm osc+}\ldots$
If the feedback is negative, the system has oscillatory solutions with
periods $T_l$ close to but slightly larger than $2\tau/(2l+1)$ for
$\varepsilon<\varepsilon^{l}_{\rm osc-}\;\;\{l=0,1,\ldots\}$, where
$0>\varepsilon^{0}_{\rm osc-}>\varepsilon^{1}_{\rm
osc-}\ldots$

Let us now discuss the bifurcation properties in the limit of large
and small time delays as well as vanishing noise and compare them with
those of a single oscillator system. The critical coupling
$\varepsilon_{\rm st}$ of the pitchfork bifurcation is time delay
independent and goes to zero for vanishing noise. However, the
critical coupling of the Hopf bifurcation depends on the time delay
(see Fig \ref{phase}, inset). As the time delay increases, the maximum
of the primary Hopf bifurcation line $\varepsilon_{\rm osc-}^1$
approaches the origin in the $(\varepsilon, D)$ plane meaning that
oscillations may occur at an arbitrary small feedback strength for the
properly tuned noise level. This should be contrasted to the dynamics
of a single noise-free oscillator with time-delayed feedback that only
exhibits oscillations at strong negative feedback $(\epsilon<-1)$.
For very small time delays $\tau\to0$, the critical coupling strength
$\varepsilon^{l,k}_{\rm osc\mp}\to\mp\infty$.



In order to compare the predictions of the dichotomous model with
the Langevin dynamics, the
peak power $P_{\rm peak}$ and the temporal mean $\langle X \rangle_t$,
resulting from the dichotomous theory, are also plotted in
Fig. \ref{mixall}. The phase diagram for the dichotomous theory is
shown in Fig. \ref{phase}.
  
Fig. \ref{mixall} and \ref{phase} show that the dichotomous theory
agrees with the Langevin dynamics quite well for small noise in the
range $D\approx0.07-0.3$ in the neighborhood of the bifurcation
points. The theory also correctly describes the bifurcation type.
Indeed, the dichotomous theory predicts accurately the noise strength
$D_{\rm H}$ ($=0.07$ for $\tau=100$) at which the Hopf bifurcation
changes from supercritical to subcritical. However, for very small $D$
the Kramers time becomes very large, and the accuracy of numerics
becomes insufficient for a comparison with the theory.

\begin{figure}
\includegraphics[width=8cm]{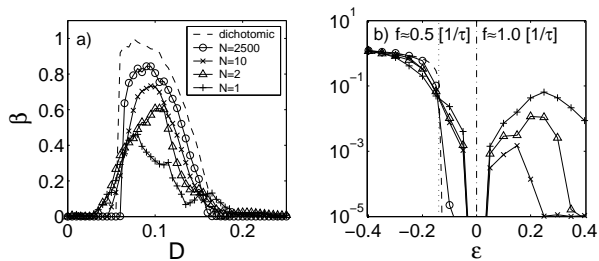}
\caption{\label{coherence} The normalized coherence of the oscillatory
states $\beta$ for systems of different size $N$ with $\tau=100$. (a)
$\beta$ as a function of the noise strength $D$ for
$\varepsilon=0.2$. (b) $\beta$ as a function of the coupling strength
$\varepsilon$ for $D=0.1$. For $\varepsilon<0$ and $\varepsilon>0$ the
spectral peak frequency is $f_p\approx0.5\;1/\tau$ and
$f_p\approx1.0\;1/\tau$, respectively. The dotted vertical line
depicts $\varepsilon^1_{\rm osc-}$.}
\end{figure}

Let us point out that the system studied in this paper exhibits the
phenomena of coherence resonance \citep[e.g.][]{Pikovsky97} and
array-enhanced resonance. Since both Kramers random switching rate $p$
(see Eq. \ref{p1221}) and the frequency of the oscillatory states
$f=\omega/(2\pi)$ (see Eq. \ref{omegaeps1}) depend on the noise
strength, i.e., $p=p(D)$ and $f=f(D)$, the noise can be tuned so that
the random hopping between the potential wells of the bistable
oscillators synchronizes with the periodic modulation of the mean
field. This statistical synchronization takes place when $f=p/2$
\citep{Gammaitoni98}, where the regularity of the oscillatory state
becomes maximal. Here, this regularity is quantified by $\beta=H
f_p/\Delta f$, where $H$ is the hight of the spectral peak at $f_p$
and $\Delta f$ is the half width of the peak.  The coherence measure
$\beta$ as a function of the noise strength is shown in
Fig. \ref{coherence}(a). We observe that the regularity of the
oscillatory states increases with increasing $N$, a property which was
reported for other systems and is sometimes referred to as
array-enhanced resonance \citep{Zhou01}. Interestingly, the
enhancement of the temporal regularity with increasing system size is
only observed for macroscopic mean field oscillations, while the
inverse holds for ``subcritical coherence''. That is, the coherence
observed in the power spectra of subcritical mean field fluctuations
(i.e., for $|\epsilon|<|\epsilon_{\rm osc\pm}|)$ decays inversely
proportional to the number of elements in the network, and becomes
negligible for $N>10$. This is shown in
Fig. \ref{coherence}(b). Qualitatively, the same dependency on the
system size is found if the delayed average does not include the
delayed element itself, i.e., the element $x_i$ is coupled to
$X_i(t-\tau)=\sum^{N-1}_{{j=1},{j\neq i}}x_j$.

In summary, we have shown that a network of noisy bistable elements
with global time-delayed coupling possesses a multiplicity of stable
oscillatory states for both positive and negative feedback in addition
to a non-zero stationary mean field for a strong enough positive
feedback which also occurs in a non-delayed system.  These novel
oscillatory states have a maximum regularity for a certain noise
strength. The bifurcations of the trivial equilibrium are well
described by the dichotomous theory in the limit of small noise and
coupling strength.  Far away from the bifurcation points the
mean-field properties of the system are well described by the Gaussian
approximation. However, the quantitative theory for the large noise
strength near the bifurcation points is still lacking. In this paper
the effect of uniform time delays on the dynamics of a globally
coupled network of bistable elements has been studied. However, many
real networks have sparse coupling and non-uniform time delays. These
properties should thus be included in a more general description.
Preliminary results of our simulations with nonuniform time-delays
suggest, that the bifurcation properties do qualitatively not change
for a wide range of Gaussian distributed time delays, which
substantiates the generic nature of the here considered model. This
issue will be discussed in detail in an upcoming article.
  
We are grateful to A.Pikovsky for many useful discussions.  This work
was supported by the Swiss National Science Foundation (D.H.) and by the
U.S. Department of Energy, Office of Basic Energy Sciences under grants
DE-FG03-95ER14516 and DE-FG-03-96ER14592 (L.T.).


\end{document}